\documentclass[11pt,a4paper,notitlepage]{article}
\usepackage[utf8]{inputenc}
\usepackage[english]{babel}
\usepackage[T1]{fontenc}
\usepackage{hyperref}
\usepackage{amsmath}
\usepackage{amsfonts}
\usepackage{color} 
\usepackage{caption}
\newcommand{\Tr}{\mathrm{Tr}}
\newcommand{\tr}{\mathrm{tr}}
\newtheorem{theorem}{Theorem}

\newtheorem{lemma}{Lemma}
\usepackage{multicol}
\usepackage{amssymb}
\usepackage{graphicx}
\usepackage[left=2.5cm,right=2.5cm,top=2.5cm,bottom=2.5cm]{geometry}
\begin{document}
\begin{center}
\textbf{\Large{Constructive Tensorial Group Field Theory \\ II:
The $U(1)-T^4_4$ Model}}
\end{center}

\begin{center}
\vspace{20pt}

Vincent Lahoche
\footnote{vincent.lahoche@th.u-psud.fr}\\
\vspace{5pt}
{\it Laboratoire de Physique Th\'eorique, CNRS-UMR 8627, Universit\'e Paris-Sud 11, 91405 Orsay Cedex, France}
\end{center}
\vspace{10pt}

\begin{abstract}
In this paper we continue our program of non-pertubative constructions of tensorial group field theories (TGFT).
We prove analyticity and Borel summability in a suitable domain of the coupling constant of
the simplest super-renormalizable TGFT which contains some ultraviolet divergencies, namely the color-symmetric
quartic melonic rank-four model with Abelian $U(1)$ gauge invariance, nicknamed $U(1)-T^4_4$.  
We use a multiscale loop vertex expansion. It is an extension of the loop vertex expansion (the
basic constructive technique for non-local theories) which is required for theories that involve non-trivial renormalization.
\end{abstract}
\tableofcontents
\pagebreak
\section{Introduction}

This paper is a sequel to \cite{LahocheLVE1}, in which we started a constructive program for tensorial group field theories (TGFT). Constructive field theory resums perturbative quantum field theory in order to obtain a rigorous definition of quantities such as Schwinger functions for interacting models \cite{rivbook}. The Loop Vertex Expansion (LVE) is a constructive technique \cite{constructivetensormodels,rivbook,LVE}, improving on the traditional constructive tools in order to treat more general models 
with non-local interactions and/or on more general geometries. Following \cite{resumgraphs}, it
 can be described as a reorganization of the perturbative series, combining an intermediate field decomposition with replicas and a forest formula. 
It allows to write the connected Schwinger functions as convergent sums indexed by spanning trees rather than as divergent sums indexed by Feynman graphs. Indeed a connected Schwinger function $S$ is usually expanded in Feynman series as
\begin{equation}
S=\sum_{\mathcal{G}}\mathcal{A}_{\mathcal{G}} ,
\end{equation}
where $\mathcal{A}_{\mathcal{G}}$ is the Feynman amplitude associated to the graph $\mathcal{G}$. However, even if each of these amplitudes are ultra-violet convergent, the sum is generally badly divergent, because of the very large number of graphs of large size, so that the perturbative expansion has a zero radius of convergence in the coupling(s), hence 
\begin{equation}
\sum_{\mathcal{G}}|\mathcal{A}_{\mathcal{G}}|=\infty  .
\end{equation}
The LVE allows to circumvent this difficulty. The first step is to consider
for any pair made of a connected graph $\mathcal{G}$ and of a spanning tree  $\mathcal{T} \subset \mathcal{G}$ in it, a universal \textit{non trivial weight} $\mathit{w}(\mathcal{G},\mathcal{T})$, which is the percentage of Hepp's sectors of $G$ in which $T$ is leading, in the sense of Kruskal \emph{greedy algorithm} (see \cite{resumgraphs} for details). These weights, being by definition \emph{percentages}, are normalized:
\begin{equation}
\sum_{\mathcal{T}\subset\mathcal{G}}\mathit{w}(\mathcal{G},\mathcal{T})=1.
\end{equation}
They allow to rewrite the Feynman expansion as a sum indexed by spanning trees:
\begin{equation}
S=\sum_{\mathcal{G}}\mathcal{A}_{\mathcal{G}}=\sum_{\mathcal{G}}\sum_{\mathcal{T}\subset\mathcal{G}}\mathit{w}(\mathcal{G},\mathcal{T})\mathcal{A}_{\mathcal{G}}=\sum_{\mathcal{T}}\mathcal{A}_{\mathcal{T}},
\end{equation}
where:
\begin{equation}
\mathcal{A}_{\mathcal{T}}:=\sum_{\mathcal{G}\supset\mathcal{T}}\mathit{w}(\mathcal{G},\mathcal{T})\mathcal{A}_{\mathcal{G}}. \label{graphsum}
\end{equation}
Since trees do not proliferate as fast as Feynman graphs, in good cases it can be shown that:
\begin{equation}
\sum_{\mathcal{T}}|\mathcal{A}_{\mathcal{T}}|<\infty  \label{treesum}
\end{equation}
at least in a certain domain that we call the \textit{summability domain}. Strictly speaking, such a program can be achieved 
with standard Feynman graphs only for Fermionic theories, because of the Pauli principle, since in that case amplitudes 
at same order combine into a determinant implying nice compensations \cite{ARfermio}. Such compensations do not occur  at fixed order 
for Bosonic theories, hence the sum \eqref{treesum} does not converge, even if it is repacked as a tree expansion. Fortunately the loop vertex expansion
overcomes this difficulty by working in another representation, called intermediate field, or Hubbard-Stratonovic. This representation
amounts to a clever exchange of the roles of propagators and vertices. 
The program summarized by equations \eqref{graphsum}-\eqref{treesum}    then works, but \emph{for
the graphs of the intermediate field representation}, and
the corresponding sum \eqref{treesum} converges absolutely to the \emph{Borel sum} of the initial expansion. \\

Group Field Theories (GFT), on the other hand, are a class of field theories defined on a group manifold and characterized by a specific form of non-locality in their interactions, giving their Feynman diagrams the structure of cellular complexes rather than graphs \cite{GFTreviews}. In some recent works, such theories appear as promising models for quantum gravity. More precisely, GFTs can be viewed either as a tentative to extend to higher dimensions the success of matrix models in dimension two \cite{Di Francesco:1993nw}, or as a second quantization of loop quantum gravity states
because spin foams arise as Feynman diagrams of GFTs \cite{GFT-LQG}.
Tensorial Group Field Theories (TGFTs) are a new class of GFTs, whose propagator is based on an inverse Laplacian \cite{GFTrenorm}
and for which interactions are chosen to be \textit{invariant}
\cite{Rivasseau-track,TGFTrenorm-Joseph,TGFTrenorm-Carrozza,TGFTrenorm-others,Lahoche:2015ola}, in the precise sense of the $U(N)^{\otimes d}$ invariance 
of rank-$d$ tensors of  size $N$ \cite{expansion1,Bonzom:2011zz,uncoloring,var-tens}. This $U(N)^{\otimes d}$ invariance provides a $1/N$ expansion. It requires tensor indices to be contracted into pairs \emph{respecting position of the indices}\footnote{This rule is the main imprvement over previous more singular tensor models \cite{tensor}.}.
The same scheme is used for TGFTs, the sums over indices being replaced by integrations over group variables. For TGFTs, it has been proved recently that this additional invariance allows to define a locality principe, the \textit{traciality}, useful for renormalization and for 
importing other classical field theory tools, such as the functional renormalization group \cite{EichhornKoslowski,BBGO,Geloun:2015qfa}.
Renormalization and phase transitions are at the core of the space-time emergence and \textit{geometrogenesis scenario} \cite{PTO}. Space-time emergence should correspond to a phase transition from a symmetric to a \textit{condensed} phase, similar to the Bose-Einstein condensation in physics of many-bodies systems. As recent works seem to confirm, \textit{asymptotic freedom} \cite{TGFTrenorm-Joseph,TGFTrenorm-others,Dine,Rivasseau-AF} phase transitions \cite{Delepouve:2015nia,Benedetti:2015ara}
 and \textit{infrared non-Gaussian fixed points} \cite{Lahoche:2015ola,BBGO,Geloun:2015qfa,Carrozza:2014rya} are common features of TGFTs, and promising steps in the long way towards understanding semi-classical space-time.\\

\medskip
The loop vertex expansion (LVE) is a technique to treat constructively non local models in the ordinary sense \cite{LVE,constructivetensormodels}. It works well for Bosonic theories without renormalization.
In case of theories requiring renormalization, such as the model of this paper, a simple LVE is not enough, and a multi-scale loop vertex expansion (MLVE)  should be used, which expands larger and larger orders of perturbation theory only when they contain 
higher and higher ultraviolet scales \cite{MLVE}. 
At least in the simplest case of super-renormalizable models,
it only requires to use \emph{two} successive forest formulas (or equivalently, a two-level \emph{jungle} formula \cite{forest}) 
instead of just one for the LVE \cite{MLVE,MLVEstandardfield,MLVEtensorfield}.

%

\medskip

This paper constructs non-perturbatively such a super-renormalizable Abelian TGFT with quartic melonic interaction at rank 4. With the MLVE technique, we prove convergence, analyticity and Borel summability theorems for the free energy and Schwinger functions of the model. Interestingly, we highlight the role of the \textit{closure constraint}, an additional invariance coming from the GFTs, advocated as a necessary ingredient for the consistent interpretation of these models as encoding simplicial geometry. This closure constraint reduces the intermediate matrix fields, turning 
them into vector fields. At the technical level, this reduction considerably simplifies the proofs.

\medskip
The plan of the paper follows the standard one of papers on this subject \cite{MLVE,MLVEtensorfield}. After recalling briefly the model and its intermediate field representation in Section 2, we introduce the standard multi-scale analysis to perform renormalization and subtract the single divergent graph of the model. In section 3, we introduce the BKAR forest-formula, and perform the MLVE in Section 4, which represent the connected functions of the theory as a two-level tree expansion that we introduce itself. This expansion contain both Bosonic and Fermionic links, and will be the first step of our proof of convergence. In Section 5, we prove the convergence theorem itself, step by step, bounding separately the Bosonic and Fermionic integrations.

\section{The Field Theory}
\subsection{A rank 4 Abelian TGFT}

One considers the Tensorial Group Field Theory (TGFT) on the Abelian group manifold $U(1)^4$, defined, in Fourier components, by the generating functional:
\begin{equation}\label{partitionfunc}
\mathcal{Z}[J,\bar{J},\lambda]=\int d\mu_{C}(T,\bar{T})e^{-S_{int}[T,\bar{T}]+\int \bar{J}T+\int \bar{T}J},
\end{equation}
with sources $J_{\vec{p}}$ and $\bar{J}_{\vec{p}}$ and the notation $\int \bar{J}T:=\sum_{\vec{p}\in\mathbb{Z}^4}\bar{J}_{\vec{p}}T_{\vec{p}}\,$; the following definition for the covariance
\begin{equation}
\int d\mu_{C}(T,\bar{T})\bar{T}_{\vec{p}^{\prime}}T_{\vec{p}}=C_{\vec{p},\vec{p}^{\,\prime}}=\delta_{\vec{p},\vec{p}^{\,\prime}}\dfrac{\delta\big(\sum_ip_i\big)}{\vec{p}^2+m^2}
\end{equation}
where the Kronecker delta $\delta\big(\sum_ip_i\big)$ implements the closure constraint \cite{TGFTrenorm-Carrozza,TGFTrenorm-Joseph}, and the action $S_{int}$ is given by:
\begin{equation}\label{int1}
S_{int}=\lambda\sum_{i=1}^4\sum_{\{\vec{p}_i\}}\mathcal{W}^{(i)}_{\vec{p_1},\vec{p_2},\vec{p}_3,\vec{p}_4}T_{\vec{p}_1}\bar{T}_{\vec{p}_2}T_{\vec{p}_3}\bar{T}_{\vec{p}_4},
\end{equation}
where the symbols $\mathcal{W}^{(i)}$ are products of Kronecker deltas:
\begin{equation}\label{bubble}
\mathcal{W}^{(i)}_{\vec{p_1},\vec{p_2},\vec{p}_3,\vec{p}_4}=\delta_{p_{1i}p_{4i}}\delta_{p_{2i}p_{3i}}\prod_{j\neq i}\delta_{p_{1j}p_{2j}}\delta_{p_{3j}p_{4j}}.
\end{equation}
Each term involved in the action $S_{int}$, with tensor indices contracted following the scheme defined by the products of Kronecker deltas given by \eqref{bubble} is called \textit{bubble}, and can be pictured graphically as a $4$-colored bipartite regular graph, with black and white vertices corresponding respectively to the fields $T$ and $\bar{T}$, joining together $4$ lines per vertex, corresponding to the Kronecker deltas. As an example, the bubble with $i=1$ is pictured in Figure \eqref{fig1}.\\
\begin{center}
\includegraphics[scale=1]{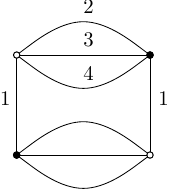} 
\captionof{figure}{Bipartite graph associated to the bubble $i=1$.}\label{fig1}
\end{center}
Schwinger - or correlation - $N$-points functions can be defined by their Feynman expansion in power of $\lambda$, indexed by \textit{Feynman graphs}:
\begin{equation}
S_N=\sum_{\{\mathcal{G}_N\}}\frac{(-\lambda)^{V(\mathcal{G}_N)}}{s(\mathcal{G}_N)}\mathcal{A}_{\mathcal{G}_N}
\end{equation}
where $\{\mathcal{G}_N\}$ is the set of graphs with $N$ external lines, $V(\mathcal{G})$ the number of vertices in $\mathcal{G}$, $s(\mathcal{G})$ a symmetry factor, and $\mathcal{A}_{\mathcal{G}_N}$ the Feynman amplitude. Any Feynman graph can be pictured graphically following the rule that bubble interaction vertices are pictured as in Figure \eqref{fig1}, and Wick contractions are pictured by dotted lines joining two black and white vertices, both in the same bubble or not. Figure \ref{fig2} gives an example of such a Feynman graph. Attributing color $0$ for the dotted lines, a Feynman graph is nothing but a bipartite regular $5$-colored graph.\\
\begin{center}
\includegraphics[scale=0.8]{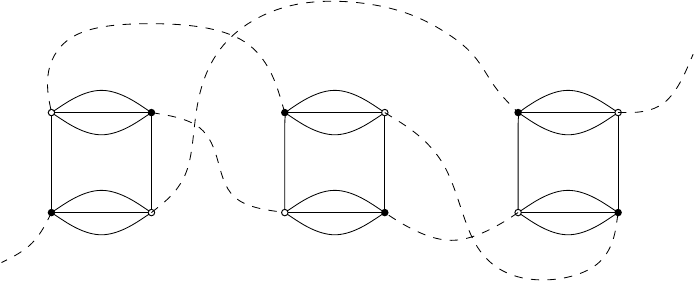}
\captionof{figure}{Example of Feynman graph with $2$ external lines and $3$ vertices}\label{fig2} 
\end{center}

\subsection{Power counting and melons}

Power counting has been established for such field theories in recent works \cite{TGFTrenorm-Carrozza,TGFTrenorm-others}, using standard multi-scale analysis, and it has been proved that the divergent degree $\omega(\mathcal{G})$ for a Feynman graph $\mathcal{G}$ is given by:
\begin{equation}
\omega(\mathcal{G})=-2L(\mathcal{G})+F(\mathcal{G})-R(\mathcal{G}).
\end{equation}
In this formula, $L(\mathcal{G})$ and $F(\mathcal{G})$ are respectively the number of lines and faces of the Feynman graph $\mathcal{G}$. We recall that a ``face" is defined as a bicolored cycle containing the color $0$, and $R(\mathcal{G})$ is the rank of the incidence matrix $\epsilon_{fe}$.\\

\noindent
The leading order contributions come from the set of \textit{melonic graphs} \cite{expansion1,expansion2,expansion3}, whose definition is recalled in \cite{LahocheLVE1}, and for which it can be shown that \cite{TGFTrenorm-Carrozza}:
\begin{equation}\label{melobound}
F(\mathcal{G})-R(\mathcal{G})=2(L(\mathcal{G})-V(\mathcal{G})+1),
\end{equation}
so that we find that the divergent degree writes as:
\begin{equation}
\omega(\mathcal{G})=2(1-V(\mathcal{G}))
\end{equation}
and the theory is \textit{super-renormalizable}. $\omega$ is negative for $V>1$. Moreover, One-vertex melonic graphs have vanishing degree and diverge logarithmically. The two possible diagrams with one vertex and two external legs are pictured in Figure \eqref{fig10}a and \eqref{fig10}b. But by direct inspection, it can be shown that the graph of Figure \ref{fig10}b is finite, with divergent degree $\,\omega=-2+1-1=-2$. The \textit{melopole}\footnote{For our purpose, a melopole is a melonic tadpole} of Figure \ref{fig10}a however has a vanishing divergent degree: $\omega=-2+3-1=0$, so that it diverges logarithmically. Hence, the only divergent non-vacuum graphs in our model requiring renormalization are melopoles\footnote{The finiteness of the number of divergent graphs is a characteristic of super-renormalizable theories.}.

\begin{center}
\includegraphics[scale=1]{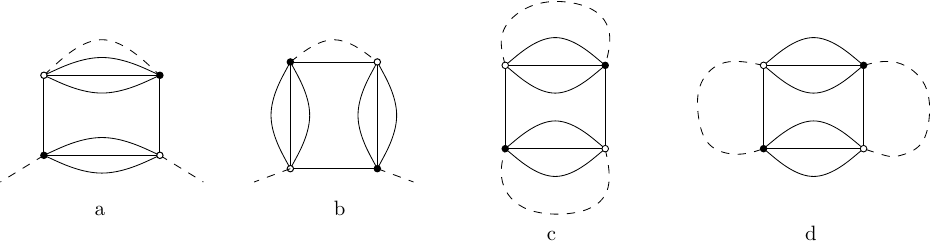} 
\captionof{figure}{The two configurations for tadpole graphs and the two vacuum graphs with one vertex}\label{fig10}
\end{center}

For a vacuum amplitude, the melonic bound \eqref{melobound} is replaced by \cite{TGFTrenorm-Carrozza}:
\begin{equation}\label{boundvacuum}
F(\mathcal{G})-R(\mathcal{G})=2(L(\mathcal{G})-V(\mathcal{G})+1)+1
\end{equation}
where the $+1$ with respect to the non-vacuum case \eqref{melobound} can be understood as follows. From a non-vacuum graph with $2N$ external lines, the closure of the graph generates $3N+1$ faces or less. A moment of reflexion shows that $R$ increases by $N$, so that the optimal variation of $F-R$ is equal to $2N+1$. At the same time, the number of internal lines increases by $N$, and the number of vertices does not change, so that the variation of $2(L-V+1)$ is equal to $2N$. With the constraint $L(\mathcal{G})=2V(\mathcal{G})-N(\mathcal{G})/2$, the divergent degree becomes:
\begin{equation}
\omega(\mathcal{G})=3-2V(\mathcal{G}).
\end{equation}
the degree is again negative for $V>1$, and the two possible configurations for $V=1$ are pictured in Figure \eqref{fig10}c and \eqref{fig10}d. A direct calculation shows that the contribution \eqref{fig10}d converges. Indeed, there are $2+3=5$ faces, $2$ lines and $R=2$, so that: $\omega=-2\times 2+(5-2)=-1$. The melonic contribution \eqref{fig10}c however, with $2\times 3+1=7$ faces is linearly divergent: $\omega=-2\times 2+(7-2)=1$, and must be renormalized. 

\subsection{Counter-terms and Renormalization}
Let us start with the non-vacuum case. Let $\mathcal{A}_{\mathcal{M}_i}(p)$ the amplitude for a melopole $\mathcal{M}_i$, of color $i$. From Feynman rules, using a sharp momentum regularization on a discrete interval $[-N,N]$, one finds:
\begin{equation}
\mathcal{A}_{\mathcal{M}_i}(p)=-2\lambda\sum_{\vec{q}\in[-N,N]^4}\frac{\delta\big(\sum_{j=1}^4q_j\big)}{\vec{q}^2+m^2}\delta_{p_iq_i}\sim \ln(N),
\end{equation}
so that only the first term, $\mathcal{A}_{\mathcal{M}_i}(0)$ of its Taylor expansion around $p=0$ diverges and must be subtracted. This subtraction can be systematically implemented with an appropriate ordering of the fields in the interaction $S_{int}$, called \textit{melordering} \cite{Carrozza1}, and consisting, for each melonic interaction bubble, in the subtraction of all the contractions over the \textit{meloforest} of the corresponding vacuum melopole. These contractions appear as \textit{mass counter-terms} in the \textit{partially renormalized classical action} $S_{int}^{PR}$, defined as:
\begin{align}
S_{int}^{PR}[\bar{T},T]&=\lambda\sum_{i=1}^4\Tr_{b_i}[\bar{T},T]+4\delta m^2\int \bar{T}T,
\end{align}
where $\Tr_{b_i}[\bar{T},T]$ is a shorthand notation for the melonic interaction labeled by $i$ and:
\begin{equation}
\delta m^2:=\mathcal{A}_{\mathcal{M}_i}(0)=-2\lambda\sum_{\vec{q}\in[-N,N]^4}\frac{\delta\big(\sum_{j=1}^4q_j\big)}{\vec{q}^2+m^2}\delta_{0q_i},
\end{equation}
so that all the non-vacuum amplitudes generated by the \textit{non-vacuum renormalized generating functional},
\begin{equation}
\int d\mu_C[\bar{T},T]e^{-S_{int}^{PR}[\bar{T},T]+\langle \bar{J},T\rangle+\langle \bar{T},J\rangle},
\end{equation}
are finite. Taking into account the vacuum divergences requires additional counter-terms, subtracting divergent graphs. From the conclusions of the previous section, the vacuum melon graphs of the type of Figure \eqref{fig10}c must be subtracted with the counter-term (one for each bubble $b_i$):
\begin{equation}
CT_v^{1}=\lambda \sum_{\vec{q}_1\in[-N,N]^4} \sum_{\vec{q}_2\in[-N,N]^4}\frac{\delta\big(\sum_jq_{1j}\big)}{\vec{q}_1^2+m^2}\frac{\delta\big(\sum_jq_{2j}\big)}{\vec{q}_2^2+m^2}\delta_{q_{11}q_{21}}.
\end{equation}
The index $1$ signals the fact that an additional counter-term is necessary to make the vacuum contributions finite. Indeed, the mass counter-term introduced previously  generates a divergent vacuum graph, and must be renormalized, with corresponding counter-term :
\begin{equation}
CT_v^{2}=-2\lambda\sum_{\vec{q}_1\in[-N,N]^4}\sum_{\vec{q}_2\in[-N,N]^4}\frac{\delta\big(\sum_jq_{1j}\big)}{\vec{q}_1^2+m^2}\frac{\delta\big(\sum_jq_{1j}\big)}{\vec{q}_2^2+m^2}\delta_{0q_{22}}.
\end{equation}
Taking into account all these counter-terms, the \textit{completely renormalized classical action}
\begin{equation}
\nonumber S_{int}^R[\bar{T},T]=\lambda\sum_{i=1}^4\Tr_{b_i}[\bar{T},T]+4\delta m^2\int \bar{T}T - 4CT_v^{1}-4CT_v^{2}
\end{equation}
subtracts all the divergences of the original model, and all the amplitudes generated by the \textit{completely renormalized generating functional}
\begin{equation}
\mathcal{Z}_{ren}[\bar{J},J]:=e^{4CT_v^{1}+4CT_v^{2}}\int d\mu_C[\bar{T},T]e^{-\lambda\sum_{i=1}^4\Tr_{b_i}[\bar{T},T]-4\delta m^2\int \bar{T}T+\langle \bar{J},T\rangle+\langle \bar{T},J\rangle}
\end{equation}
are finite. For the rest, we left the subscript "ren", but keep in mind that we deal with the renormalized generating functional. 

\subsection{Hubbard-Stratonovic decomposition}

\label{section23}Hubbard-Stratonovic (or intermediate field) decomposition is the first ingredient of the Loop Vertex Expansion.
The action \eqref{int1} with mass counter-term can be written as:
\begin{equation}\label{int2}
S_{int}^{PR}=\lambda\sum_{i=1}^4\sum_{\{\vec{p}_i\}}\mathcal{W}^{(i)}_{\vec{p_1},\vec{p_2},\vec{p}_3,\vec{p}_4}T_{\vec{p}_1}\bar{T}_{\vec{p}_2}T_{\vec{p}_3}\bar{T}_{\vec{p}_4}+4\delta m^2\sum_{\vec{p}}\bar{T}_{\vec{p}}T_{\vec{p}}.
\end{equation}
Hence, defining the three Hermitian matrices $\mathbb{M}^i$ with elements
\begin{equation}
\mathbb{M}^i_{mn}:=\sum_{\{\vec{p}_1,\vec{p}_2\}}\prod_{j\neq i}\delta_{p_{1j}p_{2j}}\delta_{p_{1i}n}\delta_{p_{2i}m}T_{\vec{p}_1}\bar{T}_{\vec{p}_2},
\end{equation}
the renormalized action \eqref{int2} can be rewritten as :
\begin{align}
S_{int}^{PR}&=\lambda\sum_{i=1}^4\bigg[\tr(\mathbb{M}^{i})^2+\frac{1}{\lambda}\delta m^2\tr(\mathbb{M}^i)\bigg]\\
&=\lambda\sum_{i=1}^4\tr\bigg[\mathbb{M}^{i}+\frac{\delta m^2}{2\lambda}\bigg]^2-(2N+1)\frac{(\delta m^2)^2}{\lambda}
\end{align}
where ``$\tr$" means the trace over indices of the matrices $\mathbb{M}^{i}$. The last term can be added to the vacuum counter-terms, so that finally the partially renormalized classical action becomes:
\begin{equation}
S_{int}^{R}=\lambda\sum_{i=1}^4\tr\bigg[\mathbb{M}^{i}+\frac{\delta m^2}{2\lambda}\bigg]^2-\bigg[(2N+1)\frac{(\delta m^2)^2}{\lambda}+CT_v^{1}+CT_v^{2}\bigg].
\end{equation}
The needed mass counter-term can then be absorbed in a global translation of the quartic interaction. The intermediate field decomposition arises as an application of the well known properties of the Gaussian integration to the partition function \eqref{partitionfunc}. Denoting$$X:=4\bigg[(2N+1)\frac{(\delta m^2)^2}{4\lambda}+CT_v^{1}+CT_v^{2}\bigg]$$ one finds:
\begin{align}\label{intermediate1}
\mathcal{Z}[J,\bar{J},\lambda]&=e^X\int d\nu_{\mathbb{I}}(\sigma)e^{-\Tr\ln(1-i\sqrt{2\lambda}C\Sigma)+i\sum_{j=1}^4\sqrt{2\lambda}\bar{\delta} m^2\tr[\sigma_j]-\bar{J}RJ}
\end{align}
where the integration over $\bar{T},T$ have been performed, $R:=(1-i\sqrt{2\lambda}C\Sigma)^{-1}C$ is the \textit{resolvent matrix}, $d\nu_{\mathbb{I}}(\sigma)$ is the normalized Gaussian integration over the $\sigma_i$ ($\mathbb{I}$ designates the covariance), $\bar{\delta}m^2:=\delta m^2/2\lambda$, and:
\begin{equation}
\Sigma:=\sum_{i=1}^4\otimes_{j=1}^{i-1}\mathbb{I}\otimes\sigma_i\otimes_{i+1}^4\mathbb{I}.
\end{equation}
The additional term $i\sum_{j=1}^4\sqrt{2\lambda}\bar{\delta} m^2\tr[\sigma_i]$ in \eqref{intermediate1} exactly compensates the divergences of the term of order $\sqrt{\lambda}$ coming from the perturbative expansion of the logarithm. As a result, the partition function \eqref{intermediate1} can be rewritten as:
\begin{equation}
\mathcal{Z}[J,\bar{J},\lambda]=e^X\int d\nu_{\mathbb{I}}(\sigma)e^{-\Tr\ln_2(1-i\sqrt{2\lambda}C\Sigma)+i\sum_{j=1}^4\sqrt{2\lambda}\sum_{p_i}A(p_i)\sigma_{i,p_ip_i}-\bar{J}RJ}
\end{equation}
with $\ln_2(1-x):=x+\ln(1-x)=O(x^2)$ and:
\begin{equation}
\sum_i\sum_{p_i}A(p_i)\sigma_{i,p_ip_i}:=\Tr(C\Sigma)+\bar{\delta}m^2\sum_i\tr(\sigma_i).
\end{equation}
Note that $-2\lambda A(p_i)$ is nothing but the renormalized amplitude $\mathcal{A}_{\mathcal{M}_i}(p_i)$ for the melopole $\mathcal{M}_i$. As a result:
\begin{equation}\label{remark}
\lambda\sum_{i=1}^4\sum_{p_i\in[-N,N]}A^2(p_i)=X.
\end{equation}

\noindent
As already explained in \cite{LahocheLVE1}, due to the closure constraint, only the \textit{diagonal part} $\tau_i(p_i):=(\sigma_i)_{p_ip_i}$ of the matrix $\sigma_i$ contributes, so that \eqref{intermediate1} writes as
\begin{align}\label{intermediatevec}
\mathcal{Z}[J,\bar{J},\lambda]=e^X\int d\nu_{\mathbb{I}}(\tau)&e^{-\sum_{\vec{p}\in\mathcal{P}}\ln_2(1-i\sqrt{2\lambda}C_0(\vec{p})\Gamma(\vec{p}))-i\sqrt{2\lambda}\sum_{j=1}^4\sum_{p_j\in\mathcal{P}}A(p_j)\tau_{j}(p_j)}\\\nonumber
&\qquad\qquad \quad\times e^{-\sum_{\vec{p}\in\mathcal{P}}\bar{J}(\vec{p})(1-i\sqrt{2\lambda}C_0(\vec{p}))\Gamma(\vec{p}))^{-1}C_{0}(\vec{p})J(\vec{p})},
\end{align}
where $C_0(\vec{p}):=(\vec{p}^2+m^2)^{-1}$, $\mathcal{P}:=\{\vec{p}\in\mathbb{Z}^4|\sum_ip_i=0\}$, $\Gamma(\vec{p}):=\sum_i\tau_i$, and $d\nu_{\mathbb{I}}(\tau)$ is the Gaussian measure of the three vector fields, defined as:
\begin{equation}
\int d\nu_{\mathbb{I}}(\tau) \tau_i(p)\tau_j(p'):=\delta_{ij}\delta_{pp'}.
\end{equation}
Interestingly, the definition \eqref{intermediatevec} can be further simplified. Indeed, because of equality \eqref{remark},
\begin{equation}\label{fieldtranslation}
\sum_{i=1}^4\sum_{p_i}\frac{1}{2}\tau_i^2(p_i)-i\sqrt{2\lambda}\sum_{j=1}^4\sum_{p_i\in\mathcal{P}}A(p_i)\tau_{i}(p_i)-X=\frac{1}{2}\sum_{i=1}^4\sum_{p_i}\big(\tau_i(p_i)-i\sqrt{2\lambda}A(p_i)\big)^2,
\end{equation}
the Gaussian measure for the intermediate field can be translated. Taking into account this translation, the partition function \eqref{intermediatevec} becomes:
\begin{align}\label{intermediatevec2}
\mathcal{Z}[J,\bar{J},\lambda]=\int d\nu_{\mathbb{I}}(\tau)&e^{-\sum_{\vec{p}\in\mathcal{P}}\ln_2(1-i\sqrt{2\lambda}C_0(\vec{p})\Gamma(\vec{p})+2\lambda D(\vec{p}))}\\
&\times e^{-\sum_{\vec{p}\in\mathcal{P}}\bar{J}(\vec{p})(1-i\sqrt{2\lambda}C_0(\vec{p}))\Gamma(\vec{p})+2\lambda D(\vec{p}))^{-1}C_{0}(\vec{p})J(\vec{p})},
\end{align}
with the definition:
\begin{equation}\label{deftrans}
D(\vec{p}):=C_0(\vec{p})\sum_iA({p}_i). 
\end{equation}
\section{BKAR Forest formula}

The BKAR (Brydges–Kennedy–Abdesselam–Rivasseau) forest interpolation formula \cite{forest}, nicknamed the ``constructive swiss knife", is the heart of the LVE. A forest formula expands a quantity defined on $n$ points in terms of forests built on these points, and is a multi-variable Taylor expansion with integral remainder. There are in fact many forest formulas, but the BKAR formula seems to be the only one which is both \textit{symmetric} under permutation of the $n$ points and \textit{positive} \cite{LVE}. \\

Let $[1, \cdots, n]$ be a finit set of points. An edge $l$ between two elements $i,j\in [1, \cdots, n]$ is a couple $(i,j)$ for $1\leq i<j\leq n$, and the set of such edges can be identified with the set of lines of $K_n$, the complete graph with $n$ vertices. Consider the vector space $S_n$ of $n\times n$ symmetric matrices, whose dimension is $n(n+1)/2$ and the \textit{compact and convex} subset $PS_n$ of \textit{positive} symmetric matrices whose diagonal coefficients are all equal to $1$, and off-diagonal elements are between $0$ and $1$. Any $X\in PS_n$ can be parametrized by $n(n-1)/2$ elements $X_l$, where $l$ run over the edges of the complete graph $K_n$ \cite{LVE}. Let us consider a smooth function $f$ defined in the interior of $PS_n$ with continuous extensions to $PS_n$ itself. The BKAR forest formula state that:
\begin{theorem}\label{BKAR} \textbf{(The BKAR forest formula)}
\begin{equation}
f(\mathbf{1})=\sum_{\mathcal{F}}\int d\mathit{w}_{\mathcal{F}}\partial_{\mathcal{F}}f[X^{\mathcal{F}}(\mathit{w}_{\mathcal{F}})]
\end{equation}
where $\mathbf{1}$ is the matrix with all entries equal to $1$, and:\\

\noindent
$\bullet$ The sum is over the forests $\mathcal{F}$ over $n$ labeled vertices, including the empty forest.\\

\noindent
$\bullet$ The integration over $d\mathit{w}_{\mathcal{F}}$ means integration from $0$ to $1$ over one parameter for each edge of the forest. Note that there are no integration for the empty forest since by convention an empty product is $1$. \\

\noindent
$\bullet$ $\partial_{\mathcal{F}}:=\prod_{l\in\mathcal{F}}\partial_l$ means a product of partial derivatives with respect to the variables $X_l$ associated to the edge $l$ of $\mathcal{F}$. \\

\noindent
$\bullet$ The matrix $X^{\mathcal{F}}(\mathit{w}_{\mathcal{F}})\in PS_n$ is such that $X^{\mathcal{F}}_{ii}(\mathit{w}_{\mathcal{F}})=1\,\forall i$, and for $i\neq j$ $X^{\mathcal{F}}_{ij}(\mathit{w}_{\mathcal{F}})$ is the infimum of the $\mathit{w}_l$ variables for $l$ in the unique path from $i$ to $j$ in $\mathcal{F}$. If no such path exists, by definition $X^{\mathcal{F}}_{ij}(\mathit{w}_{\mathcal{F}})=0$. 
\end{theorem}

\section{Multi-scale Loop Vertex Expansion}
\subsection{Slicing intermediate field decomposition}

The regularization adopted in the previous part, in the cubic domain $[-N,N]^4$ is not the most natural with respect to the rotational invariance of the Laplacian. A more natural choice, taking into account this invariance, is the restriction: $0\leq\vec{p}^2\leq N^2$. We will adopt such a cut-off for the rest of this paper. In addition, we will proceed to a \textit{slicing}, in order to make multi-scale analysis. To this end, we introduce an integer $M>1$, the ratio of a geometric progression $M^j$ so that the upper $j=j_{max}$ verifies: $M^{j_{max}}=N$, and the notation $\chi_{\leq x}(y):=\theta(x-y)$, with $\theta$ the Heaviside step function. Then, we define the following  functions on $\ell^2(\mathbb{Z}^4)$, implementing closure constraint:
\begin{align}
\chi_{\leq 1}&:=\theta(M^2-\vec{p}^2)\delta\bigg(\sum_{i=1}^4p_i\bigg)\\
\chi_{\leq j}&:=\theta(M^{2j}-\vec{p}^2)\delta\bigg(\sum_{i=1}^4p_i\bigg)\quad  j\geq 2\\
\chi_j&:=\chi_{\leq j}-\chi_{\leq j-1}\quad j\geq 2,
\end{align}
where $\chi_i$ defines the $i$-th \textit{slice}. With the definition 
\begin{equation}
U(\vec{\tau}):=i\sqrt{2\lambda}C_0(\vec{p})\Gamma(\vec{p})+2\lambda D(\vec{p})
\end{equation}
where $\vec{\tau}=(\tau_1,\tau_2,\tau_3,\tau_4)$; the interaction with cut-off $M^j$ writes
\begin{align}\label{defscalej}
V_{\leq j}&:=\Tr\ln_2(1-U_{\leq j})=\Tr[\chi_{\leq j}\ln_2(1-U)]\\
U_{\leq j}&:=i\sqrt{2\lambda}C_0(\vec{p})\Gamma(\vec{p})\chi_{\leq j}+2\lambda D(\vec{p})\chi_{\leq j},
\end{align}
and the interaction \textit{in the slice $j$} is defined as the difference:
\begin{equation}\label{Vslice}
V_j:=V_{\leq j}-V_{\leq j-1}
\end{equation}
so that the sum over scale is equal to the original interaction with cut-off $N$:
\begin{equation}
\sum_{j=0}^{j_{max}}V_j=V,
\end{equation}
and the partition function \eqref{intermediatevec2} can be written as
\begin{equation}\label{scalede}
\mathcal{Z}[J,\bar{J},\lambda]=\int d\nu_{\mathbb{I}}(\tau)\prod_{j=0}^{j_{max}}e^{-V_j}.
\end{equation}
From the definitions \eqref{defscalej} and \eqref{Vslice}, we deduce the explicit expression for $V_j$:
\begin{equation}\label{slice}
V_j:=\Tr[\chi_j\ln_2(1-U)]=\Tr\ln_2(1-U_j)
\end{equation}
with the definition:
\begin{equation}\label{definitionU}
U_{j}:=i\sqrt{2\lambda}C_0(\vec{p})\Gamma(\vec{p})\chi_{j}+2\lambda D(\vec{p})\chi_{j}.
\end{equation}

\subsection{Two-level jungle expansion}
As explained briefly in the introduction, the two-level jungle expansion which combine two successive forest-formulas play the same central role for the MLVE than the BKAR forest formula with respect to standard Loop-Vertex Expansion.\\

\noindent
To begin, we define
\begin{equation}\label{sliceweight}
W_j(\vec{\tau}):=e^{-V_j(\vec{\tau})}-1
\end{equation}
and rewrite the product over scales in \eqref{scalede} as a Grassmann integration:
\begin{equation}\label{Grassmanndec}
\mathcal{Z}[J,\bar{J},\lambda]=\int d\nu_{\mathbb{I}}(\tau)\prod_{j=0}^{j_{max}}d\mu(\bar{\eta}_j,\eta_j) e^{-\sum_j\bar{\eta}_jW_j(\vec{\tau})\eta_j}.
\end{equation}
Let $\mathcal{S}:=[0,j_{max}]$ be the integer set of scales, and $\mathbb{I}_S$ the $|\mathcal{S}|\times |\mathcal{S}|$ identity matrix, which is the covariance of the Grassmann integration measure. Hence, the previous decomposition can be rewritten as:
\begin{equation}\label{partitionexp}
\mathcal{Z}[J,\bar{J},\lambda]=\int d\nu_{\mathbb{I}}(\tau)d\mu_{\mathbb{I}_S}(\bar{\eta},\eta) e^{-W}=\sum_{n=0}^{\infty}\frac{1}{n!}\int d\nu_{\mathbb{I}}(\tau)d\mu_{\mathbb{I}_S}(\bar{\eta},\eta) (-W)^n
\end{equation}
where $W:=\sum_j\bar{\eta}_jW_j(\vec{\tau})\eta_j$, and $\eta, \bar{\eta}$ denote all the Grassmann variables collectively. The first step is to introduce a \textit{replica trick} for the Bosonic intermediate fields. We duplicate the intermediate field into copies, so that:
\begin{equation}
(-W(\tau))^n\to\prod_{m=1}^n(-W_m(\tau_m))
\end{equation}
and in the same time replace the covariance $\mathbb{I}$ by $\mathbf{1}_n$, the $n\times n$ matrix with all entries equals to $1$, so that our measure writes as $d\nu_{\mathbf{1}}(\tau_m)$. Exchanging sum and Gaussian integration, \eqref{partitionexp} becomes:
\begin{equation}\label{replica1}
\mathcal{Z}[J,\bar{J},\lambda]=\sum_{n=0}^{\infty}\frac{1}{n!}\int d\nu_{\mathbf{1}_n}(\tau_m)d\mu_{\mathbb{I}_S}(\bar{\eta},\eta) \prod_{m=1}^n(-W_m(\tau_m)).
\end{equation}
The obstacle to factorize this integral over vertices lies now in the Bosonic degenerate blocks $\mathbf{1}_n$ and the fermionic fields, which couple the vertices $W_m$. Following the method exposed in \cite{MLVE}, solving this difficulty requires two successive forest formula. The first one concerns the Bosonic fields. Introducing the \textit{coupling parameters} $x_{mp}$, so that $x_{mp}=x_{pm},\, x_{pp}=1$ between the vertex vector replicas, equation \eqref{replica1} can be rewritten, as:
\begin{equation}
\mathcal{Z}[J,\bar{J},\lambda]=\sum_{n=0}^{\infty}\frac{1}{n!}\bigg[e^{\frac{1}{2}\sum_{a,b=1}^nx_{ab}\sum_{i=1}^4\frac{\partial}{\partial\tau_i^a}\frac{\partial}{\partial \tau_i^b}+\sum_{j=0}^{j_{max}}\frac{\partial}{\partial \bar{\eta}_j}\frac{\partial}{\partial {\eta}_j}} \prod_{m=1}^n\bigg(-\sum_j\bar{\eta}_jW_{j}(\vec{\tau}_m)\eta_j\bigg)\bigg]_{\substack{\vec{\tau},\bar{\eta},\eta=0\\x_{ab}=1}}
\end{equation}
where as in  \cite{MLVE} we use the derivative formula equivalent to Gaussian integration.
Applying the BKAR forest formula for the variables $x_{ab}$, it follows:
\begin{align}
\nonumber\mathcal{Z}[J,\bar{J},\lambda]=\sum_{n=0}^{\infty}\frac{1}{n!}\sum_{\mathcal{B}_n}\int_{0}^1&\bigg(\prod_{l\in\mathcal{B}_n}d\mathit{w}_l\bigg)\bigg[e^{\frac{1}{2}\sum_{a,b=1}^nX_{ab}(\mathit{w}_l)\sum_{i=1}^4\frac{\partial}{\partial\tau_i^a}\frac{\partial}{\partial \tau_i^b}+\sum_{j=0}^{j_{max}}\frac{\partial}{\partial \bar{\eta}_j}\frac{\partial}{\partial {\eta}_j}} \\
&\times\prod_{l\in \mathcal{B}_n} \bigg(\dfrac{\partial^2}{\partial \tau_{is(l)}\partial \tau_{it(l)}}\bigg)\prod_{m=1}^n\bigg(-\sum_j\bar{\eta}_jW_{j}(\vec{\tau}_m)\eta_j\bigg)\bigg]_{\vec{\tau},\bar{\eta},\eta=0}
\end{align}
where $\mathcal{B}_n$ denotes a \textit{Bosonic forest} with $n$ vertices, and where the positive symmetric matrices $X_{ab}$ are defined in Theorem $1$ of the companion paper \cite{LahocheLVE1}. The forest $\mathcal{B}_n$ partitions the set of vertices into blocks, corresponding to its connected components, which are trees, and that we denote by $\mathfrak{V}$. Obviously, each vertex belongs to a unique Bosonic block. Contracting every Bosonic block into an ``effective vertex", we obtain a graph which we denote by $\{1,...,n\}/\mathcal{B}_n$. The last forest formula concerns Fermionic fields.  We introduce replica Fermionic fields $\eta_j^{\mathfrak{V}}$ for the effective vertices of $\{1,...,n\}/\mathcal{B}_n$, and replica coupling parameters $y_{\mathfrak{V}\mathfrak{V}'}$. Applying the forest formula to these variables, and denoting by $\mathcal{F}$ the generic Fermionic forest connecting blocks, and $\mathfrak{V}(l_f),\mathfrak{V}^\prime(l_f)$ the end blocks of the Fermionic lines in $l_f\in\mathcal{F}$, we find:
\begin{align}
\nonumber\mathcal{Z}[J,\bar{J},\lambda]&=\sum_{n=0}^{\infty}\frac{1}{n!}\sum_{\mathcal{B}_n}\sum_{\mathcal{F}}\int_{0}^1\prod_{l\in\mathcal{B}_n}d\mathit{w}_l\prod_{l_f\in\mathcal{F}}d\mathit{w}_{l_f}\\\nonumber
&\times\bigg[e^{\frac{1}{2}\sum_{a,b=1}^nX_{ab}(\mathit{w}_l)\sum_{i=1}^4\frac{\partial}{\partial\tau_i^a}\frac{\partial}{\partial \tau_i^b}+\sum_{\mathfrak{V},\mathfrak{V}'}Y_{\mathfrak{V}\mathfrak{V}'}(\mathit{w}_{l_f})\sum_{j=0}^{j_{max}}\frac{\partial}{\partial \bar{\eta}_j^{\mathfrak{V}}}\frac{\partial}{\partial {\eta}_j^{\mathfrak{V}'}}} \\\nonumber
&\times\prod_{l\in \mathcal{B}_n} \bigg(\dfrac{\partial^2}{\partial \tau_{is(l)}\partial \tau_{it(l)}}\bigg)\prod_{l_f\in\mathcal{F}}\bigg(\sum_{j=0}^{j_{max}}\big(\frac{\partial}{\partial \bar{\eta}_j^{\mathfrak{V}(l_f)}}\frac{\partial}{\partial {\eta}_j^{\mathfrak{V}'(l_f)}}+\frac{\partial}{\partial \bar{\eta}_j^{\mathfrak{V}'(l_f)}}\frac{\partial}{\partial {\eta}_j^{\mathfrak{V}(l_f)}}\big)\bigg)\\
&\times \prod_{\mathfrak{V}}\prod_{m\in\mathfrak{V}}\bigg(-\sum_j\bar{\eta}_j^{\mathfrak{V}}W_{j}(\vec{\tau}_m)\eta_j^{\mathfrak{V}}\bigg)\bigg]_{\vec{\tau},\bar{\eta},\eta=0}.
\end{align}
Note that the Fermionic lines are oriented. Expanding the sums over $j$, using the basic properties of the derivations for Bosonic and Fermionic fields, and expanding explicitly each sum over pairs of internal vertices in blocks $\mathfrak{V}$ in order to reveal the \textit{detailed Fermionic edges} $\ell_f$ between vertices in the end blocks of a given Fermionic line $l_f$ joining together these two blocks; we obtain, following  \cite{MLVE} the \textit{two-level jungle formula}:
\begin{equation}\label{twoleveljungle}
\nonumber\mathcal{Z}[J,\bar{J},\lambda]=\sum_{n=0}^{\infty}\frac{1}{n!}\sum_{\mathcal{J}}\bigg[\prod_{k=1}^n\sum_{j_k=0}^{j_{max}}\bigg]\int d\mathit{w}_{\mathcal{J}}\int d\nu_{\mathcal{J}}\partial_{\mathcal{J}}\bigg[\prod_{\mathfrak{V}}\prod_{m\in\mathfrak{V}}W_{j_m}(\vec{\tau}_m)\bar{\eta}_{j_m}^{\mathfrak{V}}\eta_{j_m}^{\mathfrak{V}}\bigg]
\end{equation}
where\\

\noindent
$\bullet$ The sum over $\mathcal{J}$ runs over all two-level jungles, hence over all oriented pairs $\mathcal{J}=(\mathcal{B}_n,\mathcal{F}_F)$ of two disjoint forests on the set $\{1,...,n\}$, such that $\bar{\mathcal{J}}=\mathcal{B}_n\cup\mathcal{F}_F$ is still a forest on $\{1,...,n\}$. The $\mathcal{B}_n$ and $\mathcal{F}_F$ are called Bosonic and fermionic components of $\mathcal{J}$. Note that the lines of $\mathcal{J}$ are partitioned into Bosonic and Fermionic lines. \\

\noindent
$\bullet$  $\int d\mathit{w}_{\mathcal{J}}$ means integration from $0$ to $1$ over parameters $\mathit{w}_{\mathcal{J}}$, one for each line in $\bar{\mathcal{J}}$, coming from forest formula. \\

\noindent
$\bullet$
\begin{equation}
\partial_{\mathcal{J}}=\prod_{l\in \mathcal{B}_n} \bigg(\dfrac{\partial^2}{\partial \tau_{is(l)}\partial \tau_{it(l)}}\bigg)\prod_{\ell_f\in\mathcal{F}_F}\delta_{j_{s(\ell_f)}j_{t(\ell_f)}}\bigg(\frac{\partial}{\partial \bar{\eta}_{j_{s(\ell_f)}}^{\mathfrak{V}(s(\ell_f))}}\frac{\partial}{\partial {\eta}_{j_{t(\ell_f)}}^{\mathfrak{V}(t(\ell_f))}}+\frac{\partial}{\partial \bar{\eta}_{j_{t(\ell_f)}}^{\mathfrak{V}(t(\ell_f))}}\frac{\partial}{\partial {\eta}_{j_{s(\ell_f)}}^{\mathfrak{V}(s(\ell_f))}}\bigg)
\end{equation}
where $\mathfrak{V}(m)$ denotes the Bosonic blocks to which $m$ belongs. \\

\noindent
$\bullet$ The measure $d\nu_{\mathcal{J}}$, mixing Bosonic and Fermionic integrations is defined as, for some $F$:
\begin{equation}
\int d\nu_{\mathcal{J}} F:=e^{\frac{1}{2}\sum_{a,b=1}^nX_{ab}(\mathit{w}_l)\sum_{i=1}^4\frac{\partial}{\partial\tau_i^a}\frac{\partial}{\partial \tau_i^b}+\sum_{\mathfrak{V},\mathfrak{V}'}Y_{\mathfrak{V}\mathfrak{V}'}(\mathit{w}_{l_f})\sum_{m\in\mathfrak{V},m^{\prime}\in\mathfrak{V}}\delta_{j_mj_{m^{\prime}}}\frac{\partial}{\partial \bar{\eta}_{j_{m}}^{\mathfrak{V}}}\frac{\partial}{\partial {\eta}_{j_{m^{\prime}}}^{\mathfrak{V}'}}}F\bigg|_{\vec{\tau},\bar{\eta},\eta=0}
\end{equation}

\noindent
Since the slice assignments, the fields, the measure and the integrand are now factorized over the connected components of $\bar{\mathcal{J}}$, the logarithm of $Z$ is easily computed as the restriction of the previous sum \eqref{twoleveljungle} to the two-level spanning trees (the connected component of the two-level forests):
\begin{equation}\label{exp1}
\ln\mathcal{Z}[J,\bar{J},\lambda]=\sum_{n=1}^{\infty}\frac{1}{n!}\sum_{\mathcal{J}\,tree}\bigg[\prod_{k=1}^n\sum_{j_k=0}^{j_{max}}\bigg]\int d\mathit{w}_{\mathcal{J}}\int d\nu_{\mathcal{J}}\partial_{\mathcal{J}}\bigg[\prod_{\mathfrak{V}}\prod_{m\in\mathfrak{V}}W_{j_m}(\vec{\tau}_m)\bar{\eta}_{j_m}^{\mathfrak{V}}\eta_{j_m}^{\mathfrak{V}}\bigg].
\end{equation}

\noindent
The rest of this paper is devoted to the proof of the following Theorem\\

\begin{theorem}\label{keythm}
Let $\lambda=\rho e^{i\phi}$, $\phi\in(-\pi,\pi)$. For $\rho$ small enough, the series \eqref{exp1} is absolutely and uniformly convergent in $j_{max}$, for $g$ in the small open cardioid domain defined by $|\lambda|\leq \rho \cos (\phi/2)$. The ultra-violet limit $\ln(\mathcal{Z})=\lim_{j_{max}\to\infty}\ln(\mathcal{Z}[\lambda,j_{max}])$ is therefore well-defined and analytic in that cardioid domain, and is the Borel sum of its perturbative expansion in powers of $\lambda$. 
\end{theorem}

\section{Bounds and convergence}
\subsection{The Grassmann integrals}

The\label{sectionboundgrass} sum \eqref{exp1} splits into Grassmann and Bosonic integrals, and we start with the first. As explained in \cite{MLVE}, due to the standard properties of Grassmann integration, the Gaussian integration over these variables can be written as:
\begin{align}
\nonumber\prod_{\mathfrak{V}}\prod_{m\in\mathfrak{V}}&\bigg(\frac{\partial}{\partial \bar{\eta}_{j_m}^{\mathfrak{V}}}\frac{\partial}{\partial \eta_{j_m}^{\mathfrak{V}}}\bigg)e^{\sum_{\mathfrak{V},\mathfrak{V}'}Y_{\mathfrak{V}\mathfrak{V}'}(\mathit{w}_{l_f})\sum_{m\in\mathfrak{V},m^{\prime}\in\mathfrak{V}}\delta_{j_mj_{m^{\prime}}}\bar{\eta}_{j_{m}}^{\mathfrak{V}}{\eta}_{j_{m^{\prime}}}^{\mathfrak{V}'}}\\
&\times\prod_{\ell_f\in\mathcal{F}_F}\delta_{j_{s(\ell_f)}j_{t(\ell_f)}}\bigg(\bar{\eta}_{j_{s(\ell_f)}}^{\mathfrak{V}(s(\ell_f))}{\eta}_{j_{t(\ell_f)}}^{\mathfrak{V}(t(\ell_f))}+\bar{\eta}_{j_{t(\ell_f)}}^{\mathfrak{V}(t(\ell_f))}{\eta}_{j_{s(\ell_f)}}^{\mathfrak{V}(s(\ell_f))}\bigg)\bigg|_{\bar{\eta},\eta=0}.
\end{align}
Denoting $\textbf{Y}_{mm^{\prime}}:=Y_{\mathfrak{V}(m)\mathfrak{V}(m^{\prime}})\delta_{j_mj_{m^{\prime}}}$, and taking into account that this matrix is symmetric, the previous Gaussian integral turns to the more familiar form:
\begin{equation}\label{inttrans}
\int \prod_{\mathfrak{V}}\prod_{m\in\mathfrak{V}}d\bar{\eta}_{j_m}^{\mathfrak{V}}d\eta_{j_m}^{\mathfrak{V}} e^{-\sum_{m,m^\prime}^n\bar{\eta}_{j_m}^{\mathfrak{V}(m)}\textbf{Y}_{mm^{\prime}}\eta_{j_m}^{\mathfrak{V}(m^{\prime})}}\prod_{\ell_f\in\mathcal{F}_F}\delta_{j_{s(\ell_f)}j_{t(\ell_f)}}\bigg(\bar{\eta}_{j_{s(\ell_f)}}^{\mathfrak{V}(s(\ell_f))}{\eta}_{j_{t(\ell_f)}}^{\mathfrak{V}(t(\ell_f))}+\bar{\eta}_{j_{t(\ell_f)}}^{\mathfrak{V}(t(\ell_f))}{\eta}_{j_{s(\ell_f)}}^{\mathfrak{V}(s(\ell_f))}\bigg).
\end{equation}
Defining:
\begin{equation}\label{minor}
\textbf{Y}_{m_1,...,m_k}^{p_1,...,p_k}:=\int \prod_{\mathfrak{V}}\prod_{m\in\mathfrak{V}}d\bar{\eta}_{j_m}^{\mathfrak{V}}d\eta_{j_m}^{\mathfrak{V}} e^{-\sum_{m,m^\prime}\bar{\eta}_{j_m}^{\mathfrak{V}(m)}\textbf{Y}_{mm^{\prime}}\eta_{j_{m^\prime}}^{\mathfrak{V}(m^{\prime})}}\prod_{r=1}^k{\eta}_{m_{r}}^{\mathfrak{V}(r)}\bar{\eta}_{p_{r}}^{\mathfrak{V}(r)},
\end{equation}
and taking into account what the authors of  \cite{MLVE} have called the \textit{hard core constraint inside each blocks}, meaning that the integral \eqref{inttrans} vanishes if two vertices belong to the same Bosonic block $\mathfrak{V}$ with the same scale attribution, they write \eqref{inttrans} as:
\begin{equation}
\bigg(\prod_{\mathfrak{V}}\prod\limits_{\substack{m,m^{\prime}\in\mathfrak{V}\\m\neq m^{\prime}}}(1-\delta_{j_{m}j_{m^{\prime}}})\bigg)\bigg(\prod_{\ell_f\in\mathcal{F}_F}\delta_{j_{s(\ell_f)}j_{t(\ell_f)}}\bigg)\bigg(\textbf{Y}_{m_1,...,m_k}^{p_1,...,p_k}+\textbf{Y}_{p_1,...,m_k}^{m_1,...,p_k}+\cdots+\textbf{Y}_{p_1,...,p_k}^{m_1,...,m_k}\bigg)
\end{equation}
where the sum runs over the $2^k$ ways to exchange the upper and lower indices, and $k:=|\mathcal{F}_F|$ is the cardinal of the Fermionic forest, and the first product implements the hard core constraint. For our purpose, the following result, for which a proof can be found in  \cite{MLVE}, is relevant to achieve the fermionic bound:
\begin{lemma}
Due to the positivity of the covariance $\textbf{Y}$, for any $\{m_i\}$ and $\{p_i\}$ the minor $\textbf{Y}_{m_1,...,m_k}^{p_1,...,p_k}$ defined in \eqref{minor} satisfies:
\begin{equation}
|\textbf{Y}_{m_1,...,m_k}^{p_1,...,p_k}|\leq 1.
\end{equation}
\end{lemma}

\subsection{Bosonic integrals}

We now move on to the problem of the Bosonic integrals, whose bound is more subtle than the Fermionic one. From formula \eqref{exp1}, Bosonic integration factorizes over each blocks $\mathfrak{V}$. As a result, we can only consider and bound one of these block contributions. Let 
us consider such a block $\mathfrak{V}$. It involves the Gaussian integration:
\begin{equation}\label{bosonicint}
\int d\nu_{\mathfrak{V}} F_{\mathfrak{V}}(\vec{\tau})=e^{\frac{1}{2}\sum_{a,b\in\mathbb{B}}X_{ab}(\mathit{w}_l)\sum_{i=1}^4\frac{\partial}{\partial\tau_{i,a}}\frac{\partial}{\partial \tau_{i,b}}}F_{\mathfrak{V}}(\vec{\tau})\big|_{\vec{\tau}=0}
\end{equation}
with $F_{\mathfrak{V}}(\vec{\tau})$ defined as:
\begin{equation}\label{defF}
F_{\mathfrak{V}}(\vec{\tau})=\prod_{l\in \mathfrak{V}} \bigg(\dfrac{\partial^2}{\partial \tau_{i,s(l)}\partial \tau_{i,t(l)}}\bigg)\prod_{m\in\mathfrak{V}}W_{j_m}(\vec{\tau}_m)\,,
\end{equation}
where, as from the beginning, the sums over the momenta being implied. The derivatives $\partial/\partial \tau$ can be evaluated from the famous Faà di Bruno formula, extending the standard derivation rule for composed functions, and easily proved by induction:
\begin{equation}
\partial_x^{q}f(g(x))=\sum_{\pi}f^{|\pi|}(g(x))\prod_{B\in\pi}g^{|B|}(x),
\end{equation}
where $\pi$ runs over the partitions of the set $\{1,...,q\}$ and $B$ runs through the blocks of the partition $\pi$. With this helpful result, and from the definitions \eqref{slice} and \eqref{sliceweight}, we have:
\begin{equation}
\partial_{\tau_i}(-V_j)=-i\sqrt{2\lambda}\sum_{\vec{p}\in\mathcal{P}_i}C_0(\vec{p})\chi_j(1-R_j(\vec{p}))=i\sqrt{2\lambda}\sum_{\vec{p}\in\mathcal{P}_{\bot i}}C_0(\vec{p})\chi_jU_jR_j(\vec{p}),
\end{equation}
where $\mathcal{P}_{\bot i}$ is the subset of $\mathcal{P}$ where the component $p_i$ is frozen, and equals to the external momentum. This formula can be easily extended for a derivative of degree $k>0$ as:
\begin{equation}
\prod_{l=1}^k\partial_{\tau_{i(l)}}(-V_j)=(i\sqrt{2\lambda})^k(k-1)!\left(\sum_{\vec{p}\in\mathcal{P}_{j,\bot{\{k\}}}}C_0^k(\vec{p})R_j^{k} \right)
\end{equation}
where $\mathcal{P}_{j,\bot{\{k\}}}$ is the intersection of the support of the distribution $\chi_j$ on the slice $j$ with the gauge invariant subset $\mathcal{P}_{\bot{\{k\}}}$, frozen each of the external momenta to their external values, and setting all the external momenta with the same color equals together. Moreover:
\begin{equation}
R_j:=\frac{1}{1-U_j}.
\end{equation}
The $k$-th derivative of $W_j$ can be deduced from Faà di Bruno formula. For $k>0$:
\begin{equation}
\prod_{l=1}^k\partial_{\tau_{i(l)}}(-W_j)=e^{-V_j}\sum\limits_{\substack{\{m_l\}\\\sum_{l\geq1}lm_l=k}}\frac{k!}{\prod_{l\geq1}m_l!(l!)^{m_l}}\prod_{l\geq 1}[\partial_{\tau}^l(-V_j)]^{m_l}\,,
\end{equation}
where the compact notation $\prod_{l\geq 1}[\partial_{\tau}^l(-V_j)]^{m_l}$ means partitions of size $m_l$ of the original product $\prod_{l=1}^k\partial_{\tau_{i(l)}}$. In \eqref{defF}, we can rewrite the product as a product over the \textit{arcs} of the vertices: 
\begin{equation}\label{defF2}
F_{\mathfrak{V}}(\vec{\tau})=\prod_{m\in\mathfrak{V}}\prod_{k=1}^{c(m)} \dfrac{\partial}{\partial \tau_{i(k),a(m)}}W_{j_m}(\vec{\tau}_m).
\end{equation}
where $c(m)$ is the \textit{coordination number} of the vertex $m$, equal to the number of half lines of the intermediate-fields hooked to this vertex. Then, the Bosonic integral \eqref{bosonicint} becomes:
\begin{align}\label{gaussianformula}
\int d\nu_{\mathfrak{V}}\bigg[&\prod_{m\in\mathfrak{V}}e^{-V_j}(i\sqrt{2\lambda})^{c(m)}\sum\limits_{\substack{\{x_l^{(m)}\}\\\sum_{l\geq1}lx_l^{(m)}=c(m)}}\frac{c(m)!}{\prod_{l\geq1}x_l^{(m)}!l^{x_l^{(m)}}} \overline{\prod_{l\geq 1}[\partial_{\tau}^l(-V_j)]^{x_l(m)}}\bigg]\,,
\end{align}
where, once again we used of the compact notation $\prod_{l\geq 1}[\partial_{\tau}^l(-V_j)]^{x_l(m)}$, the bar meaning we extracted the factor $(i\sqrt{2\lambda})^{l\times x_l^{(m)}}$. Note that in this formula the sums over momenta along the intermediate field edges is implicit as well. 
In order to bound the resolvant insertions, we have the following helpful lemma (see \cite{MLVEtensorfield})
\begin{lemma}\label{boundresolvent}
Because $C_0$ and $\Gamma$ are real, and $D(\vec{p})$ is positive, for $\rho$ small enough $R_{j_m}$ obeys the following bounds:
\begin{equation}
|R_j|\leq 2\cos^{-1}\left(\frac{\phi}{2}\right)\,,
\end{equation}
with $\phi:=\arg(\lambda)\in ]-\pi ,\pi ]$. 
\end{lemma}
Then, using the constraint: $\sum_m c(m)=2(|\mathfrak{V}|-1)$, with $|\mathfrak{V}|$ the number of vertices of $\mathfrak{V}$, \eqref{gaussianformula} admits the bound:
\begin{align}
\nonumber\big|\int d\nu_{\mathfrak{V}} F_{\mathfrak{V}}(\vec{\tau})\big|\leq &\bigg\vert\frac{8\lambda}{\cos^2(\phi/2)}\bigg\vert^{|\mathfrak{V}|-1} \int d\nu_{\mathfrak{V}}\bigg[\prod_{m\in\mathfrak{V}}e^{-V_{j_m}(\vec{\tau}_m)}\\
&\sum\limits_{\substack{\{x_l^{(m)}\}\\\sum_{l\geq1}lx_l^{(m)}=c(m)}}\frac{c(m)!}{\prod_{l\geq1}x_l^{(m)}!l^{x_l^{(m)}}} \overline{\prod_{l\geq 1}\vert\partial_{\tau}^l(-V_j)\vert^{x_l(m)}_{R=\mathbb{I}}}\bigg]\,,
\end{align}
the notation $R=\mathbb{I}$ meaning that all the resolvant are setting equal to the identity operator. Moreover, note that $U_{j_m}$ involves intermediate fields. Then, defining:
\begin{equation}
G_{\mathfrak{V}}:=\prod_{m\in\mathfrak{V}}\sum\limits_{\substack{\{x_l^{(m)}\}\\\sum_{l\geq1}lx_l^{(m)}=c(m)}}\frac{c(m)!}{\prod_{l\geq1}x_l^{(m)}!l^{x_l^{(m)}}}
\overline{\prod_{l\geq 1}\vert\partial_{\tau}^l(-V_j)\vert^{x_l(m)}_{R=\mathbb{I}}}\,,
\end{equation}
and since the Gaussian measure $d\nu_{\mathfrak{V}}$ is positive, we can use the Cauchy-Schwarz inequality to get:
\begin{equation}\label{CSbound}
\int d\nu_{\mathfrak{V}} \prod_{m\in\mathfrak{V}}e^{-V_{j_m}(\vec{\tau}_m)}G_{\mathfrak{V}}\leq \bigg(\int d\nu_{\mathfrak{V}}\prod_{m\in\mathfrak{V}}\big|e^{-2V_{j_m}(\vec{\tau}_m)}\big|\bigg)^{1/2}\bigg(\int d\nu_{\mathfrak{V}} |G_{\mathfrak{V}}|^2\bigg)^{1/2}.
\end{equation}
We shall treat separately each term, calling the first term the \textit{non-perturbative factor}, and the second the \textit{perturbative factor}, following the conventions of  \cite{MLVE}. Note that in our derivation of the bound \eqref{CSbound}, we have not considered the special case for which the tree has one vertex only. This particular contribution involves melonic vacuum diagrams, discarded by construction with their corresponding counter-terms, and non-melonic vacuum diagrams. Because these diagrams are convergent, these contributions can be easily bounded, and they do not spoil the conclusion\footnote{This can be easily proved rigorously with integration by part with respect to the intermediate field, following a standard strategy exposed for instance in  \cite{MLVE}.}.

\subsection{Bound of the Bosonic integral}

We\label{sectionboundgauss} begin with the first term, the non perturbative contribution:
\begin{equation}
B_1:=\int d\nu_{\mathfrak{V}}\prod_{m\in\mathfrak{V}}\bigg|e^{-2V_{j_m}(\vec{\tau}_m)}\bigg|.
\end{equation}
Firstly, note that: $\big|e^{-2V_{j_m}(\vec{\tau}_m)}\big|\leq e^{2|V_{j_m}(\vec{\tau}_m)|}$. Secondly, because of the identity:
\begin{equation}
\ln_2(1-x)=\int _0^1dt \frac{tx^2}{1-tx},
\end{equation}
we have, from Lemma \eqref{boundresolvent}:
\begin{equation}
|V_j|\leq \frac{2}{\cos(\phi/2)}\bigg|\sum_{\vec{p}}U_j(\vec{p})\bigg|^2\leq \frac{2}{\cos(\phi/2)}\sum_{\vec{p}}|U_j(\vec{p})|^2
\end{equation}
and we get:
\begin{equation}
B_1\leq \int d\nu_{\mathfrak{V}}\prod_{m\in\mathfrak{V}}\exp\bigg(\frac{4}{\cos(\phi/2)}\sum_{\vec{p}}|U_{j_m}(\vec{p}\,)|^2\bigg).
\end{equation}
Using Definition \eqref{intermediatevec2},
\begin{equation}
\sum_{\vec{p}}|U_j(\vec{p})|^2=2|\lambda|\big[C_0^2(\vec{p})\Gamma^2(\vec{p})+2|\lambda|D^2(\vec{p})+2\sqrt{2|\lambda|}C_0(\vec{p})D(\vec{p})\Gamma(\vec{p})\big].
\end{equation}
From Definition \eqref{deftrans} of $D(\vec{p})$, and because the renormalized function $A(p)$ behaves as $\ln(p^2+m^2)$, $D^2(\vec{p})\leq \mathcal{O}(1)$. Similarly, $\sum_{\vec{p}}\delta_{p_1p}C_{0}(\vec{p})D(\vec{p})\leq \mathcal{O}(1)$. For $\lambda$ small enough, we deduce that:
\begin{equation}
\sum_{\vec{p}}|U_j(\vec{p})|^2\leq 2\lambda \big[\mathcal{O}(1)+\sum_{\vec{p}\in\mathcal{P}_j}C_{0}^2(\vec{p})\Gamma^2(\vec{p})\big]\,.
\end{equation}
Now, note that $C_{0}^2(\vec{p})\chi_j(\vec{p})\leq \mathcal{O}(1) M^{-4j}\chi_j(\vec{p})$, therefore:
\begin{equation}
\sum_{\vec{p}}C_0^2(\vec{p})\chi_j(\vec{p})\Gamma^2(\vec{p})=\sum_{i,j}\sum_{p_i,p_j}L(p_i,p_j)\tau_{im}(p_{i})\tau_{jm}(p_{j})\,,
\end{equation}
with the definition:
\begin{equation}
L(p_1,p_2):=\mathcal{O}(1)M^{-4j}\sum_{p_3,p_4}\chi_j(\vec{p})\delta(p_1+p_2+p_3+p_4)\,.
\end{equation}
As a result:
\begin{equation}
\sum_{\vec{p}}|U_{j_m}(\vec{p})|^2\leq 2\lambda \sup\big(\mathcal{O}(1)\big)\bigg[1+\sum_{i,j}\sum_{p_{i},p_{j}}L(p_i,p_j)\tau_{im}(p_{i})\tau_{jm}(p_{j})\bigg]
\end{equation}
where the notation $\sup\big(\mathcal{O}(1)\big)$ stands for the highest of the numerical constants involved in the first bound. The term of degree $2$ in $\tau$ gives an effective variance $X^{-1}_{\mathfrak{V}}\delta_{p_i,p_j}-\bigg|\frac{16\lambda}{\cos(\phi/2)}\bigg|\mathbb{I}_{\mathfrak{V}}L(p_i,p_j)$, where $X_{\mathfrak{V}}$ is the covariance of the Gaussian measure and $\mathbb{I}_{\mathfrak{V}}$ is the identity matrix in replica space. The Gaussian integration can be  computed, and gives (taking into account the normalization of original Gaussian measure):
\begin{equation}
B_1\leq e^{\mathcal{O}(1)\big|\frac{8\lambda|\mathfrak{V}|}{\cos(\phi/2)}\big|}\times \det\bigg[\mathbf{1}-\bigg|\frac{16\lambda}{\cos(\phi/2)}\bigg|L \otimes X_{\mathfrak{V}}\bigg]^{-1/2}\,,
\end{equation}
where $\mathbf{1}$ denote the identity matrix in the whole space on which the determinant is computed. The determinant can be computed in terms of traces with the formula $\det(1-X)=e^{\Tr\ln(1-X)}$. Denoting:
\begin{equation}
X=\bigg|\frac{16\lambda}{\cos(\phi/2)}\bigg| L \otimes X_{\mathfrak{V}},
\end{equation}
we have:
\begin{equation}
\Tr(X)=\mathcal{O}(1) M^{-j}\bigg|\frac{64\lambda}{\cos(\phi/2)}\bigg||\mathfrak{V}|\leq \mathcal{O}(1)\sum_j M^{-j}\bigg|\frac{64\lambda}{\cos(\phi/2)}\bigg||\mathfrak{V}|\leq \mathcal{O}(1) \bigg|\frac{64\lambda}{\cos(\phi/2)}\bigg||\mathfrak{V}|,
\end{equation}
and, for the norm of $X$:
\begin{equation}
||X||\leq \bigg|\frac{64\lambda}{\cos(\phi/2)}\bigg|.
\end{equation}
where in both cases we used the fact that all diagonal entries of $X_{\mathfrak{V}}$ are equal to $1$. Finally, using the Taylor expansion $-\ln(1-X)=\sum_{n\geq1}X^n/n$, the two previous bounds imply:
\begin{equation}
-\Tr\ln(1-X)=\sum_{n\geq1}\frac{\Tr(X^n)}{n}\leq \Tr(X)\sum_{n\geq2}\frac{||X||^n}{n}\leq|\mathfrak{V}|\times \sum_{n\geq 1}\bigg|\frac{64\lambda}{\cos(\phi/2)}\bigg|^n
\end{equation}
and for $\lambda$ small enough, we find:
\begin{equation}\label{boundb1}
B_1\leq e^{\mathcal{O}(1)\big|\frac{2\lambda}{\cos(\phi/2)}\big||\mathfrak{V}|}.
\end{equation}

We now move on to the perturbative bound:
\begin{equation}\label{perturbative}
B_2:=\bigg(\int d\nu_{\mathfrak{V}} |G_{\mathfrak{V}}|^2\bigg)^{1/2}.
\end{equation}
For $l>1$ we have:
\begin{equation}
\sum_{\vec{p}\in\mathcal{P}_j}C_0^l(\vec{p})\leq \sum_{\vec{p}\in\mathcal{P}_j}C_0^2(\vec{p})\leq \frac{1}{M^{4(j-1)}} \sum_{\vec{p}\in\mathcal{P}_j}1.
\end{equation}
The last sum can be bounded by the integral over the volume of the intersection between the plane of $\mathbb{R}^4$ of equation $\sum_{i=1}^4p_i=0$ and the volume in between the hyper-spheres of equations $\vec{p}^2=M^{2(j-1)}$ and $\vec{p}^2=M^{2j}$. This volume corresponds to the volume between the two spheres of $\mathbb{R}^3$ of radius $M^j$ and $M^{j-1}$, times a factor $1/2$ coming from the normalization of Kronecker delta:
\begin{equation}
\sum_{\vec{p}\in\mathcal{P}_j}1\leq \frac{1}{2}\times \big(\mathbb{V}_j-\mathbb{V}_{j-1}\big)\leq \frac{2}{3}\pi M^{3j}
\end{equation}
and for the worst case:
\begin{equation}
\sum_{\vec{p}\in\mathcal{P}_j}C_0^l(\vec{p})\leq \dfrac{2}{3}\pi M^4 M^{-j}.
\end{equation}
The Gaussian integrals can be computed more easily by reversing the field translation \eqref{fieldtranslation}. Because, obviously: $|e^{-i\sqrt{2\lambda}\sum_{j=1}^4\sum_{p_j\in\mathcal{P}}A(p_j)\tau_{j}(p_j)}|\leq 1$, we can treat the integral for the back-translated intermediate fields with the simple replacement: $C_0(\vec{p})U_{j_m}(\vec{p})\to i\sqrt{2\lambda}C_0(\vec{p})\chi_{j_m}\Gamma_m(\vec{p})$, and an additional factor $e^{-4\Tr(X_{\mathcal{B}})\sum_{p}A^2(p)\chi_j}\leq 1$. In the worst case, the Gaussian integration is bounded by the following one:
\begin{equation}\label{gaussianexemple}
H_{\mathfrak{V}}:=\int d\nu_{\mathfrak{V}}\prod_{m\in\mathfrak{V}}\bigg(\sum_{\vec{p}_{\alpha_1}\in\mathcal{P}_{j_m}}C_0^2(\vec{p})\Gamma_m(\vec{p})\bigg)^{k_m}.
\end{equation}
Such an integral can be pictured as a graph with $|\mathfrak{V}|$ vertices, labeled by $m$, each of them having $k_m$ half colored intermediate field lines hooked to them. By Wick theorem, the Gaussian integration joins together half lines between the vertices. In the worst case the graph has no loop, and because $||X_{\mathcal{B}}||\leq 1$, it follows that the Gaussian integration \eqref{gaussianexemple} is bounded by:
\begin{align}
|H_{\mathfrak{V}}|\leq 4^{\sum_mk_m}\prod_{m\in\mathfrak{V}}\bigg(\dfrac{2}{3}\pi M^4 M^{-{j_m}}\bigg)^{k_m}\times \bigg(\sum_{m\in\mathfrak{V}}k_m\bigg)!!
\end{align}
where we have taken into account the $4$ choices for the colors of intermediate field lines through the factor $4^{\sum_mk_m}$. For \eqref{perturbative}, the double factorial is:
\begin{equation}
\bigg(2\sum_{m\in\mathfrak{V}}x_1^{(m)}\bigg)!!\leq (4|\mathfrak{V}|-4)!!
\end{equation}
Note that the combinatorial factor:
\begin{equation}\label{rq}
\sum\limits_{\substack{\{x_l^{(m)}\}\\\sum_{l\geq1}lx_l^{(m)}=c(m)}}\frac{1}{\prod_{l\geq1}x_l^{(m)}!l^{x_l^{(m)}}}
\end{equation}
is nothing but the coefficient of $x^{c(m)}$ in the Taylor expansion of $\prod_{k}e^{x^k/k}=1/(1-x)$. We find the bound:
\begin{equation}
|B_2|\leq \sqrt{(4|\mathfrak{V}|-4)!!}\times \prod_{m\in\mathfrak{V}}\sum\limits_{\substack{\{x_l^{(m)}\}\\\sum_{l\geq1}lx_l^{(m)}=c(m)}}\frac{c(m)!}{\prod_{l\geq1}x_l^{(m)}!l^{x_l^{(m)}}}\vert 2\lambda\vert^{x_1^{(m)}/2}\bigg(\frac{8\pi}{3}\bigg)^{c(m)}M^{-(j_m-4)}
\end{equation}
where, following  \cite{MLVE}, we assume that the scales $j$ have an inferior bound $j_{min}>4$ (this is certainly not essential, since the first slices
can be treated by a simple LVE). Choosing $|\lambda|\leq 1/2$, and with the remark following \eqref{rq}, we find finally the pessimistic bound
\begin{equation}\label{boundb2}
|B_2|\leq \sqrt{(4|\mathfrak{V}|-4)!!}\times \prod_{m\in\mathfrak{V}}\bigg(\frac{8\pi}{3}\bigg)^{c(m)}c(m)!M^{-j_m}.
\end{equation}
Taking into account the bound \eqref{boundb1}, we find, using \eqref{CSbound}:
\begin{align}
\big|\int d\nu_{\mathfrak{V}} &\prod_{m\in\mathfrak{V}}e^{-V_{j_m}(\vec{\tau}_m)}G_{\mathfrak{V}}\big|
\leq e^{\mathcal{O}(1)\big|\frac{2\lambda}{\cos(\phi/2)}\big||\mathfrak{V}|}\sqrt{(4|\mathfrak{V}|-4)!!}\times \prod_{m\in\mathfrak{V}}c(m)!\bigg(\frac{8\pi}{3}\bigg)^{c(m)}M^{-j_m}
\end{align}
and:
\begin{align}
\big|\int &d\nu_{\mathfrak{V}} F_{\mathfrak{V}}(\vec{\tau})\big|\leq e^{\mathcal{O}(1)\big|\frac{2\lambda}{\cos(\phi/2)}\big||\mathfrak{V}|}\bigg(\frac{8\lambda}{\cos^2(\phi/2)}\bigg)^{|\mathfrak{V}|-1}\sqrt{(4|\mathfrak{V}|-4)!!}\times \prod_{m\in\mathfrak{V}}c(m)!\bigg(\frac{8\pi}{3}\bigg)^{c(m)}M^{-j_m}.
\end{align}

\subsection{Final Bound}

Collecting together the results of sections \eqref{sectionboundgrass} and \eqref{sectionboundgauss}, we find the bound for the expansion \eqref{exp1} of the free energy $\ln(Z)$ (we omit the exponential factor to simplify the expressions):
\begin{align}\label{firststep}
\nonumber|\ln\mathcal{Z}[J,\bar{J},\lambda]|&\leq\sum_{n=1}^{\infty}\frac{1}{n!}\sum_{\mathcal{J}\,tree}\bigg[\prod_{k=1}^n\sum_{j_k=0}^{j_{max}}\bigg]2^{L(\mathcal{F}_F)}\bigg(\prod_{\ell_f\in\mathcal{F}_F}\delta_{j_{s(\ell_f)}j_{t(\ell_f)}}\bigg)\prod_{\mathfrak{V}}\prod\limits_{\substack{m,m^{\prime}\in\mathfrak{V}\\m\neq m^{\prime}}}(1-\delta_{j_{m}j_{m^{\prime}}})\\
&\times\bigg(\frac{8|\lambda|}{\cos^2(\phi/2)}\bigg)^{|\mathfrak{V}|-1}\sqrt{(4|\mathfrak{V}|-4)!!}\times \prod_{m\in\mathfrak{V}}c(m)!\bigg(\frac{8\pi}{3}\bigg)^{c(m)}M^{-j_m}\,,
\end{align}
where $L(\mathcal{F}_F)$ denote the number of fermionic lines.  At this stage, the reasoning follows exactly that of  \cite{MLVE}. Thanks to \textit{Cayley's Theorem}, the number of trees with $n$ labeled vertices and coordination numbers $c_i$ for each vertex $i=1,...,n$ is given by:
\begin{equation}\label{Cayley}
\frac{(n-2)!}{\prod_i(c_i-1)!}.
\end{equation}
This result shows that the  sum involved in \eqref{firststep} obeys
\begin{equation}
\sum\limits_{\substack{c(m)|\sum_mc(m)=2\\\mathfrak{V}|-2}} \prod_{m\in\mathfrak{V}}c(m)=\frac{(3|\mathfrak{V}|-3)!}{(|\mathfrak{V}|-2)!(2|\mathfrak{V}|-1)!}\; .
\end{equation}
Collecting all the factorials leads to:
\begin{equation}
\sqrt{(4|\mathfrak{V}|-4)!!}\frac{(3|\mathfrak{V}|-3)!}{(2|\mathfrak{V}|-1)!}.
\end{equation}
Using  Stirling's formula as in  \cite{MLVE}
\begin{equation}\label{boundcomb}
2\sqrt{(4|\mathfrak{V}|-4)!!}\frac{(3|\mathfrak{V}|-3)!}{(2|\mathfrak{V}|-1)!}\leq (|\mathfrak{V}|-1)! 3^{|3\mathfrak{V}|}e^{-|\mathfrak{V}|}|\mathfrak{V}|^{|\mathfrak{V}|}.
\end{equation}
We now move on to sum over scale attributions, taking into account the hard core constraint. As explained in details in  \cite{MLVE}, the hard core constraint imposes that the scale assignments of vertices in a same block are all different, which implies:
\begin{equation}
\sum_{m\in\mathfrak{V}}j_m\geq j_{min}+(j_{min}+1)+\cdots +(j_{min}+|\mathfrak{V}|-1)=j_{min}|\mathfrak{V}|+\frac{|\mathfrak{V}|(|\mathfrak{V}|-1)}{4}
\end{equation}
and:
\begin{equation}
\sum_{m\in\mathfrak{V}}(j_m-2)\geq \frac{1}{2}\sum_{m\in\mathfrak{V}}j_m+\frac{j_{min}-4}{2}|\mathfrak{V}|+\frac{|\mathfrak{V}|(|\mathfrak{V}|-1)}{4},
\end{equation}
where we have introduced explicitly the minimal scale $j_{min}>4$. This result implies that, 
\begin{align}
\sum_{\{j_m\}}\prod\limits_{\substack{m,m^{\prime}\in\mathfrak{V}\\m\neq m^{\prime}}}(1-\delta_{j_{m}j_{m^{\prime}}})\prod_{m\in\mathfrak{V}}M^{-j_m}\leq \bigg(\sum_{j=j_{min}}^{j_{max}}M^{-j/2}\bigg)^{|\mathfrak{V}|}\frac{1}{M^{\frac{j_{min}-4}{2}|\mathfrak{V}|+\frac{|\mathfrak{V}|(|\mathfrak{V}|-1)}{4}}}
\end{align}
which, for $j_{min}>4$ and $M>4$, is uniformly bounded by $M^{-|\mathfrak{V}|^2/4}$. The upper bound $j_{max}$ can now be sent to infinity without any divergence, allowing to define non-perturbatively the ultraviolet limit of the theory. \\

The final step is to sum over the Fermionic forest. Such a forest can be partitioned into components of cardinal $b_k$, associated to connected blocks of size $k$, hence with $k$ sub-vertices. The number of Fermionic lines is then $\sum_kb_k-1$. For each component with $k$ sub-vertices, there are $n^{b_k}$ ways to hook a Fermionic line. Moreover, from \cite{MLVE}, the number of uncolored two-level trees is bounded by $4^n n^{n-2}$. Taking into account the color leads to $4^{2n} n^{n-2}$. We find:
\begin{align}
\nonumber|\ln\mathcal{Z}[J,\bar{J},\lambda]|&\leq \sum_{n}\frac{4^n}{n!}\sum\limits_{\substack{\{b_k\}\\\sum_kkb_k=n}}\frac{n!}{\prod_kb_k!(k!)^{b_k}}2^{\sum_kb_k-1}k^{(\sum_kb_k)-2}\prod_kn^{b_k}\\
&\times\prod_k\bigg[\bigg(\frac{512\pi^2|\lambda|}{9\cos^2(\phi/2)}\bigg)^{k-1}\sqrt{(4k-4)!!}\frac{(3k-3)!}{(2k-1)!}M^{-k^2/4}\bigg]^{b_k} .
\end{align}
Taking into account the bound \eqref{boundcomb} as well as the constraint $\sum_{k}kb_k=n$, when the number of (sub) vertices is fixed to $n$, and the easy bound coming from Stirling formula : $n^{(\sum_kb_k)-2}\leq (\sum_kb_k)!e^n$, we find the final bound:
\begin{align}\label{finalstep}
|\ln\mathcal{Z}[J,\bar{J},\lambda]|&\leq \sum_{b\geq 0}\bigg[\sum_{n\geq1}\bigg(\frac{512\pi^2|\lambda|}{9\cos^2(\phi/2)}\bigg)^{n-1}(4\times 3^{3})^nn^nM^{-n^2/4}\bigg]^b.
\end{align}
The power of  $M$ ensures that, for $M$ sufficiently large, this factor compensates the bad divergence associated to $n^n$. The radius of convergence is then finite, and the factor $\cos^2(\phi/2)$ establishes the domain of uniform convergence as stated in Theorem \eqref{keythm}. 

\section{Conclusion}
In this first constructive paper for super-renormalizable TGFTs, we have successfully applied the Multi-scale Loop Vertex Expansion to the simplest super-renormalizable TGFT. The theorem of this paper can easily be extended to connected Schwinger functions, introducing additional resolvents for each pair of derivations with respect to the external sources. Interestingly, the fact that the closure constraint reduces the intermediate fields' freedom degrees to vector-like intermediate fields considerably simplified the proof of the convergence. In particular, in \cite{MLVEtensorfield} some technical difficulties occurring from the non-commutativity of the operators involved in the resolvents forced us to use complicated iterated Cauchy-Schwarz estimates. They disappear 
in our TGFT thanks to the closure constraint. \\

\noindent
The next step of this constructive program would be to construct the same model for $d=5$ and $d=6$. Presumably, the extension to the $d=5$ case should be doable, because it is still super-renormalizable, and requires only of a finite number of subtractions. The just-renormalizable $d=6$ case, however, may be more difficult. However, a promising indication for the future is that this theory is ultra-violet asymptotically free. 

\section{Acknowledgments}
The author specially thanks his supervisor Vincent Rivasseau for his suggestions at the initiative of this project, his advice and his patient proofreading; and his wife Julie Lahoche for the languages corrections.

\end{document}